\pdfoutput=1
\documentclass[a4paper,american,citeautoscript,floatfix,pdftex,superscriptaddress,twoside,%
aps,prl,%
longbibliography,%
reprint,twocolumn,
]{revtex4-1}%
\usepackage{amsfonts,amsmath,amssymb}
\usepackage[T1]{fontenc}
\usepackage{graphicx}%
\usepackage[utf8]{inputenc}
\usepackage{microtype}
\usepackage[obeyFinal,textsize=footnotesize]{todonotes}
\usepackage{xspace}
\usepackage{hyperref, hypernat}
\usepackage[todos]{CMImacros}

\graphicspath{{./figure/}} 
\hypersetup{citebordercolor=yellow,linkbordercolor=red,urlbordercolor=blue}

\renewcommand{\figureautorefname}{Fig.}%
\renewcommand{\tableautorefname}{Table}%

\newcommand{\suppinf}{Suppl.\ Inf.\xspace}%
\makeatletter%
\renewcommand*{\@fnsymbol}[1]{\ensuremath{\ifcase#1\or \|\or \mathsection\or \ddagger\or *\or **\or
      \mathparagraph\or \dagger\or \dagger\dagger \or \ddagger\ddagger \else\@ctrerr\fi}}%
\makeatother

\newcommand{\cfeldesy}{\affiliation{Center for Free-Electron Laser Science, Deutsches
      Elektronen-Synchrotron DESY, Notkestraße 85, 22607 Hamburg, Germany}}%
\newcommand{\uhhcui}{\affiliation{Center for Ultrafast Imaging, Universität Hamburg, Luruper
      Chaussee 149, 22761 Hamburg, Germany}}%
\newcommand{\uhhphys}{\affiliation{Department of Physics, Universität Hamburg, Luruper Chaussee 149,
      22761 Hamburg, Germany}}%
\newcommand*{\jilinu}{\altaffiliation[Permanent address: ]{Institute of Atomic and Molecular
      Physics, Jilin University, Changchun 130012, China}}%
\newcommand*{\unibasel}{\altaffiliation[Present address: ]{Department of Chemistry, University of
      Basel, Klingelbergstrasse 80, 4056 Basel, Switzerland}}%
\newcommand*{\uninijmegen}{\altaffiliation[Present address: ]{Institute for Molecules and Materials,
      Radboud University, Heijendaalseweg 135, 6525 AJ Nijmegen, The Netherlands}}%
\newcommand{\jkemail}{\email[Corresponding author: ]{jochen.kuepper@cfel.de}}%
\newcommand{\cmiweb}{\homepage{https://www.controlled-molecule-imaging.org}}%

\begin{document}
\title{Pure Molecular Beam of Water Dimer}%
\author{Helen Bieker}\cfeldesy\uhhphys\uhhcui%
\author{Jolijn Onvlee}\cfeldesy%
\author{Melby Johny}\cfeldesy\uhhphys%
\author{Lanhai He}\jilinu\cfeldesy%
\author{Thomas~Kierspel}\unibasel\cfeldesy\uhhphys\uhhcui%
\author{Sebastian Trippel}\cfeldesy\uhhcui%
\author{Daniel A. Horke}\uninijmegen\cfeldesy\uhhcui%
\author{Jochen Küpper}\jkemail\cmiweb\cfeldesy\uhhphys\uhhcui%

\begin{abstract}\noindent%
   Spatial separation of water dimers from water monomers and larger water-clusters through the
   electric deflector is presented. A beam of water dimers with $93~\%$ purity and a rotational
   temperature of $1.5~$K was obtained. Following strong-field ionization using a 35~fs laser pulse
   with a wavelength centered around 800~nm and a peak intensity of $10^{14}~\Wpcmcm$ we observed
   proton transfer and $46~\%$ of ionized water dimers broke apart into hydronium ions \HHHOp and
   neutral OH.
\end{abstract}
\maketitle


\section{Introduction}
Hydrogen bonding between water molecules plays an important role in aqueous systems, \eg, for
biomolecules that are surrounded by solvents. It is responsible for the unique properties of water,
such as its high boiling point~\cite{Jeffrey:HydrogenBonding:1997}. While hydrogen bonds have been
studied extensively in many different molecular systems \cite{Liu:Science271:929,
   Nauta:Science287:293, Dunning:Science347:530, Berden:JCP104:972, Korter:JPCA102:7211,
   Sobolewski:JPCA40:9275, Ren:NatPhys:79:1745}, one of the most important models remains the water
dimer, somehow the smallest drop of water. Numerous studies have been conducted on this benchmark
system and its structure with a single hydrogen bond is well known~\cite{Odutola:JCP72:1980,
   Yu:JCP121:2004, Dyke:JCP66:1977, Dyke:JCP57:5011}.

Water molecules and water-clusters have been studied using various techniques to describe dynamics
such as proton motion~\cite{Marchenko:PhysRevA98:063403} or chemical processes, \eg, reactive
collisions~\cite{Kilaj:NatComm9:2096}. For investigations of ultrafast molecular dynamics, such as
energy and charge transfer across hydrogen bonds in molecular systems, photoion-photoion coincidence
measurements at free-electron lasers are developing as a powerful tool~\cite{Boll:SD3:043207,
   Kierspel:Dissertation:2016, Ren:NatPhys:79:1745} and this approach was also used to study
hydrogen bonding in the water dimer at a synchrotron~\cite{Jahnke:NatPhys6:139}. Other spectroscopic
techniques utilizing synchrotron facilities \cite{Winter:JCP126:124504, Smith:Science306:851} or
table-top laser-systems~\cite{Keutsch:PNAS98:10533, Berden:JCP104:972, Korter:JPCA102:7211,
   Zwier:ARPC47:205} further improved the knowledge about hydrogen bonding in water and
water-clusters.

Most of these experiments investigating the dynamics of hydrogen-bonded systems would benefit from
samples of identical molecules in a well-defined initial state. The widely used supersonic expansion
technique provides cold molecular beams down to rotational temperatures of
$<1$~K~\cite{Scoles:MolBeam:1, Even:JCP112:8068, Johny:CPL721:149}. However, cluster expansions do
not produce single-species beams, but a mixture of various cluster stoichiometries. Hence, only low
concentrations of specific species can be achieved. In the case of water molecules, supersonic
expansion produces a cold beam of various water clusters~\cite{Liu:Science271:929} with a water
dimer concentration of only a few percent~\cite{Jahnke:NatPhys6:139, Paul:JPCA101:5211}. This leads
to small experimental event rates and requires long measurement times, \eg, in coincidence detection
schemes~\cite{Jahnke:NatPhys6:139, Kierspel:Dissertation:2016}. These experiments with a mixture of
molecules in a molecular beam are only feasible if it can be disentangled which molecule was
actually measured. Therefore, these mixtures severely limit the applicable techniques. A pure beam
of water dimers would significantly speed up the measurements, when unwanted backgrounds from
carrier gas and larger water-clusters are avoided, or simply enable such experiments.

The electrostatic deflector is an established method to spatially separate the molecules of interest
from the carrier gas and to separate different species within a cold molecular
beam~\cite{Chang:IRPC34:557}. This includes the separation of molecular
conformers~\cite{Filsinger:PRL100:133003, Filsinger:ACIE48:6900, Kierspel:CPL591:130,
   Teschmit:ACIE57:13775}, individual quantum states of small molecules~\cite{Nielsen:PCCP13:18971,
   Horke:ACIE53:11965}, as well as specific molecular clusters~\cite{Trippel:PRA86:033202,
   You:JPCA122:1194, Johny:CPL721:149}. The deflector was previously utilized in investigations of
water, \eg, to determine the rotational temperatures of ``warm'' molecular beams of
water~\cite{Moro:PRA75:013415}, to separate its \emph{para} and \emph{ortho}
species~\cite{Horke:ACIE53:11965}, and to measure the dipole moment of small
water-clusters~\cite{Moro:PRL97:123401}. Alternatively, separation by the cluster species' distinct
collision cross sections, \ie, by the transverse momentum changes due to scattering with a
perpendicular rare-gas beam, was demonstrated~\cite{Buck:CR100:3863}; this method is especially
amenable to larger cluster sizes~\cite{Pradzynski:Science337:1529}. Such spatially separated
single-species samples enable, for instance, advanced imaging applications of water-clusters using
non-species-specific techniques, as well as the study of size-specific effects and the transition
from single-molecule to bulk behavior.

\section{Experimental Methods}
Here, the electrostatic deflector was used to spatially separate water dimers from water monomers as
well as larger water-clusters in a molecular beam formed by supersonic expansion. The experimental
setup was described previously~\cite{Chang:IRPC34:557, Trippel:RSI89:096110}. Briefly, liquid water
was placed in the reservoir of an Even-Lavie valve~\cite{Even:JCP112:8068}, heated to \celsius{55},
seeded in 100~bar of helium, and expanded into vacuum with a nominal driving-pulse duration of
19.5~\us and at a repetition rate of 250~Hz. The produced molecular beam was doubly skimmed, 6.5~cm
($\varnothing=3$~mm) and 30.2~cm ($\varnothing=1.5$~mm) downstream from the nozzle, directed through
the electrostatic deflector~\cite{Kienitz:JCP147:024304} of 154~mm length and with a nominal field
strength of 50~\kVpcm with an applied voltage of 8~kV across the deflector, before passing through a
third skimmer ($\varnothing=1.5$~mm). The deflector was placed 4.4~cm behind the tip of the second
skimmer. In the center of a time-of-flight (TOF) mass spectrometer, 134.5~cm downstream from the
nozzle, molecules were strong-field ionized by a 35~fs short laser pulse with a central wavelength
around 800~nm and a pulse energy of 170~\uJ. Focusing to 65~\um yielded a peak intensity of
$\ordsim10^{14}~\Wpcmcm$. The generated ions were accelerated toward a microchannel-plate detector
combined with a phosphor screen and the generated signal was recorded with a digitizer. The valve,
skimmers, and deflector were placed on motorized translation stages, which allowed movement of the
molecular beam through the ionization laser focus and the recording of vertical
molecular-beam-density profiles without moving the laser focus, resulting in fixed imaging
conditions~\cite{Kuepper:PRL112:083002, Stern:FD171:393, Filsinger:JCP131:064309}.

While the employed strong-field ionization is a general, non-species specific ionization technique,
it can also lead to fragmentation of molecules, such that recorded mass spectra (MS) do not directly
reflect the composition of the molecular beam. In combination with the species-specific deflection
process, however, this can be disentangled and, thus, even allows for the investigation of
strong-field-induced fragmentation processes of a single species.

\section{Results and Discussion}
\begin{figure}
   \includegraphics[width=0.5\textwidth]{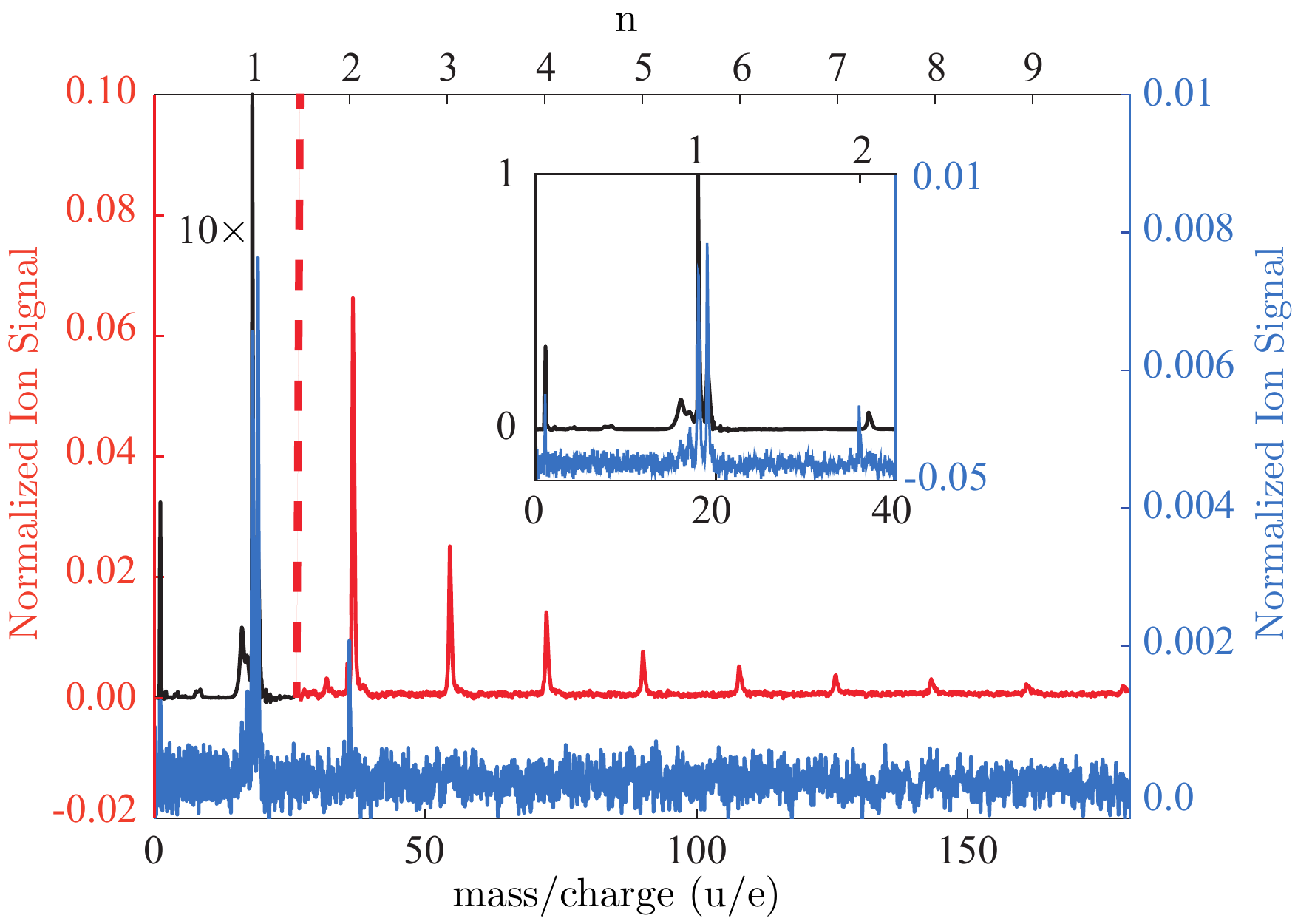}
   \caption{TOF-MS in the center of the molecular beam as depicted in \autoref{fig:deflection} with
      a deflector voltage of 0~kV (black, red) and at a position of $+3$~mm with a deflector voltage
      of 8~kV (blue). For mass/charge ($m/q$) ratios of $0\ldots30$~u/e, to the left of the dashed
      red line, the TOF-MS has been scaled by $0.1$. The inset shows the region of
      $m/q=0\ldots40$~u/e enlarged; see text for details.}
   \label{fig:TOF}
\end{figure}
TOF-MS of the direct and the deflected beams are shown in \autoref{fig:TOF}. The spectrum of the
undeflected beam shows water-cluster ions $(\HHO)_n^+$ up to $n=2$ and protonated water-cluster ions
$(\HHO)_n\text{H}^+$ up to $n=10$. Even larger clusters were likely formed in the supersonic
expansion, but were not observed in the recorded TOF interval. We point out that all clusters that
reach the interaction region are neutral clusters of the type $(\HHO)_{n}$, and protonated clusters
must result from the interactions with the femtosecond laser, \ie, due to fragmentation during or
after the strong-field-ionization process.

\begin{figure}
   \includegraphics[width=0.5\textwidth]{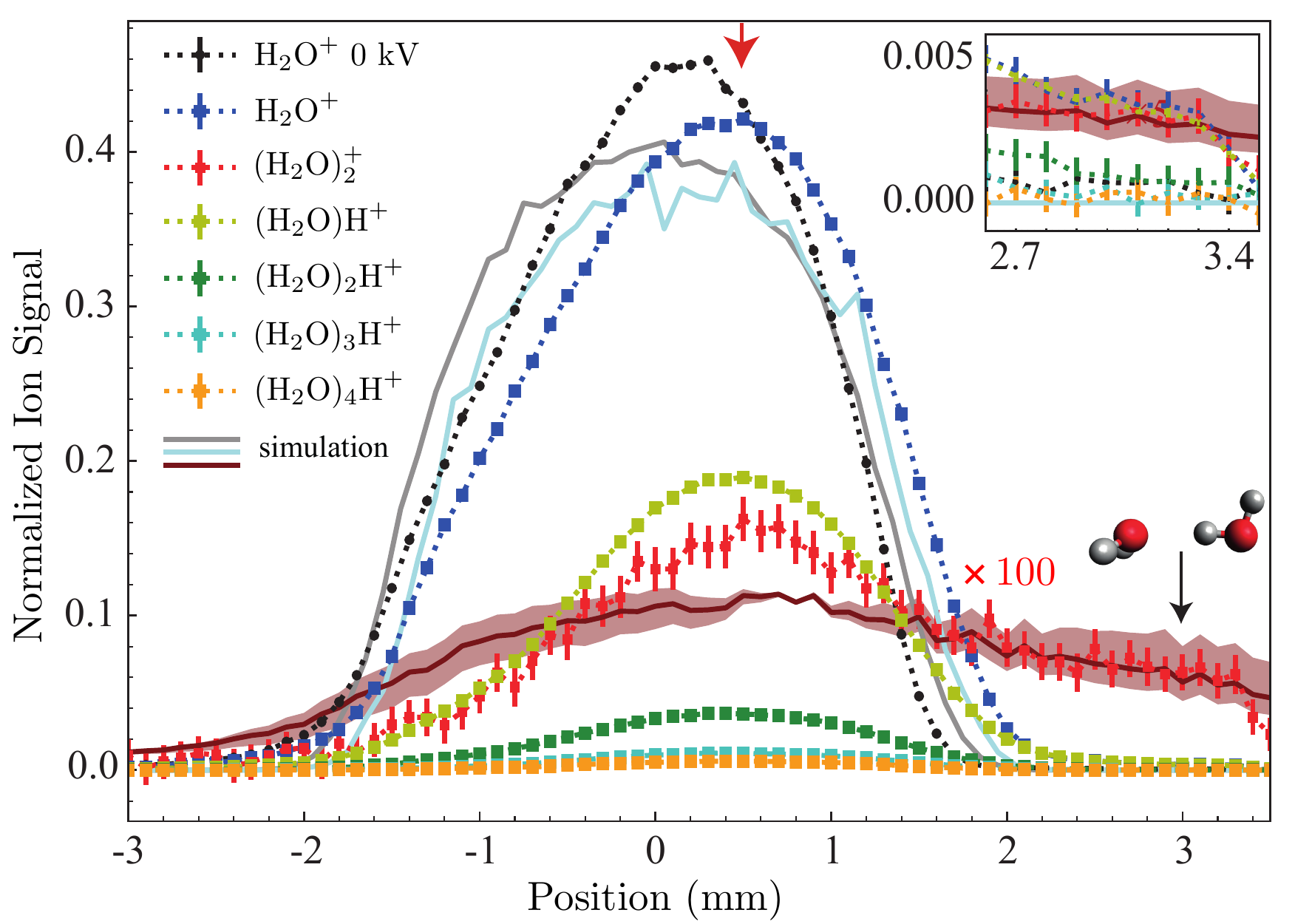}
   \caption{Normalized measured vertical molecular-beam-density profiles (dashed lines) of water
      cation $(\HHO)^+$ (blue), water-dimer cation $(\HHO)_{2}^+$ (red), and protonated
      water-cluster ions $(\HHO)_n$H$^+$ up to $n=4$ (yellow, green, cyan, orange) with deflector
      voltages of 0~kV (black circles) and 8~kV (squares). Simulated vertical molecular-beam-density
      profiles of the undeflected water monomer (grey) as well as of the deflected water monomer
      (light blue) and water dimer (dark red) with a deflector voltage of 8~kV are shown as solid
      lines. The shaded (dark red) area depicts the error estimate of the water dimer simulation due
      to the temperature uncertainty, $T_{\text{rot}}=1.5(5)$~K; see text for details. The black
      arrow indicates the position in the deflected beam where the TOF-MS shown in \autoref{fig:TOF}
      was measured. The inset shows the deflection region enlarged with a magnification factor of
      $5$ applied to the $(\HHO)_{2}^+$ and $(\HHO)_n$H$^+$ signals.}
   \label{fig:deflection}
\end{figure}
Vertical molecular-beam-density profiles for water ions $(\HHO)^+$, water dimer ions $(\HHO)_{2}^+$,
and protonated water-cluster ions $(\HHO)_n$H$^+$ up to $n=4$, with a potential difference of 8~kV
applied across the deflector, are shown in \autoref{fig:deflection}. For comparison, a field-free
vertical profile for the water ion with 0~kV across the deflector is also shown. The vertical
molecular-beam-density profiles have been normalized to the area of the field-free spatial profile
of the water ion. For visibility the water dimer profile has been scaled by a factor of 100 after
normalization. While the field-free molecular beam profile is centered around 0~mm, application of a
voltage of 8~kV to the deflector shifted the peak of water ions, water dimer ions, and protonated
water-cluster ions by +0.5~mm, as indicated by the red arrow in \autoref{fig:deflection}. In
addition, water dimer ions showed a broadening and an increase of signal at around +3~mm, indicated
by a black arrow in \autoref{fig:deflection}.

In the inset of \autoref{fig:deflection} the region around $+3$~mm is shown enlarged with a
magnification factor of 5 applied to $(\HHO)_{2}^+$ and $(\HHO)_n$H$^+$ with $n=1\ldots4$. The
corresponding TOF-MS in the deflected part of the beam at a position of +3~mm is highlighted in
\autoref{fig:TOF} by the blue line. Not just water dimer ions, but also hydronium ions, \HHHOp, and
water ions, \HHOp, showed an increased signal in the deflected beam. The shape of the vertical beam
profiles for these ions matched the water dimer profile in the region of 2.8--3.5~mm, indicating
that they originated from the same parent molecule.

The water dimer ion was the largest non-protonated cluster measured in this setup. To verify that
the water dimer ion was originating from the water dimer, the deflection behaviour of water-clusters
inside the electrostatic deflector was simulated. Therefore, the Stark energies and effective dipole
moments of water monomers and water-clusters were calculated with the freely available
\textsc{CMIstark} software package~\cite{Chang:CPC185:339} using rotational constants, dipole
moments, and centrifugal distortion constants from the literature~\cite{DeLucia:JPCR3:211,
   Coudert:JMolSpec139:259, Shostak:JCP94:5875, Malomuzh:RussJPhysChemA88:1431, Dyke:JCP66:1977},
see \suppinf \tableautorefname~I; contributions of the polarizability to the Stark effect could
safely be ignored~\cite{Maroulis:JCP113:1813, Kienitz:JCP147:024304}. The rotational constants of
the water dimer are significantly smaller than for the water monomer, leading to a larger effective
dipole moment for the water dimer than for water and a larger acceleration in the electric field in
the deflector, see \figureautorefname~3 and \tableautorefname~I of the \suppinf for further
information.

The simulated vertical molecular-beam-density profiles of the water monomer and the water dimer are
shown in \autoref{fig:deflection}. The deviations between the measured and simulated undeflected
vertical beam profiles are ascribed to imperfect alignment of experimental setup, which was not
taken into account in the simulations. Due to the rotational-state dependence of the Stark effect,
the deflection of a molecular beam in an electrostatic field depends on the rotational temperature
of the molecular ensemble~\cite{Chang:IRPC34:557} and the best fit for the profiles of the water
monomer and the water dimer at a deflector voltage of 8~kV was obtained assuming a Boltzmann
population distribution of rotational states corresponding to $1.5(5)$~K.

Not only deflection of water-clusters measured as a mass of 36~amu, but also of water-clusters
detected as protonated-clusters have been measured, for instance, for $(\HHO)_2^+$, as indicated by
the red arrow and symbols in \autoref{fig:deflection}. Trajectory simulations for water-clusters
$(\HHO)_n^+$ with $n=3\ldots7$ using a rotational temperature of $T_\text{rot}=1.5(5)$~K were
performed to understand the origin of this deflection behavior. For the water hexamer three and for
the water heptamer two conformers have been simulated, assuming an equal population of the
conformers. These showed that, based on the different effective dipole moments, a different
deflection is expected for different water-clusters, see \figureautorefname~5 of the \suppinf. Since
the detected protonated water-clusters arose from the strong-field fragmentation of larger neutral
clusters in the interaction region, the measured vertical protonated-cluster density profiles are a
superposition of several neutral water-cluster density profiles. Thus, it is not possible to compare
the individual simulated molecular-beam-density profiles of neutral clusters directly with the
measured protonated water-cluster density profiles. Instead, at each position of the deflection
profile the signal from all water-clusters has been summed up, both for the simulated and the
measured molecular-beam-density profiles. The result yields a comparable amount of deflection for
simulated and measured molecular-beam-density profiles, see \suppinf \figureautorefname~6. The shift
of 0.5~mm can, therefore, originate from the superimposed molecular-beam-density profiles from
different larger clusters due to fragmentation into smaller water-clusters. The same shift is
visible for $\HHO^+$ and $(\HHO)_2^+$, which indicates that water-clusters are also fragmenting into
$\HHO^+$ and $(\HHO)_2^+$. Nevertheless, the simulation for water-clusters $n=1\ldots7$ shows that
the water dimer deflected the most, reaching a position of +3~mm and above, see \suppinf
\figureautorefname~4 and \figureautorefname~5. Of all the other clusters considered, only the water
hexamer in its prism and book forms reaches to a position up to 3.2~mm with the falling edge of the
profile. In our experiments the water hexamer and higher order clusters have only been measured as
fragments, such that the concentration and size distribution of neutral clusters in the molecular
beam is unknown. However, the measured fragment distributions strongly suggest that significantly
larger clusters are not present, since the ion signals decay exponentially and it is known that
clusters primarily fragment through loss of single water molecules~\cite{Angel:CPL345:2001,
   Belau:JPCA111:2007, Liu:cpl4:270}.

The TOF-MS in the deflected part of the beam, shown in \autoref{fig:TOF}, contains peaks
corresponding to H$^{+}$, O$^{+}$, OH$^{+}$, $\HHOp$, and $\HHHOp$, in addition to the water dimer
ion. As mentioned before the short-pulse ionization can lead to fragmentation. For the water dimer,
two fragmentation channels were reported for electron-impact ionization with 70~eV
electrons~\cite{Angel:CPL345:2001}: either an $\HHHOp$ ion and a neutral OH are formed or a $\HHO^+$
ion and a neutral water monomer \HHO. Using a size-selection method and infrared spectroscopy,
$\HHO^+$ has been reported as a fragment of the water dimer~\cite{Buck:CR100:3863}. Comparison of
the vertical molecular-beam-density profiles of the deflected molecules allowed further
investigation of the fragmentation channels of the water dimer. The measured vertical
molecular-beam-density profiles of these molecules showed a similar deflection behavior in the
region of 2.8 to 3.5~mm as the water dimer, see \autoref{fig:deflection} and \suppinf
\figureautorefname~1. The observed constant ratio of those fragments over this spatial region
indicates that all these fragments originated from the water dimer.

Comparing the intensity of the fragments of the water dimer, $\HHO^{+}$ and $\HHHOp$ and
($\HHO)_2^+$, in the deflected beam, the fragmentation ratios of the intact water dimer were
estimated. These showed that $46(7)~\%$ of the water dimer fragmented into one ionized water
molecule, while $46(4)~\%$ of the water dimer underwent most likely proton transfer and formed a
hydronium ion. Only $8(2)~\%$ of the water dimer present in the molecular beam stayed intact after
ionization.

The actual number of water dimer molecules per shot in the deflected molecular beam was estimated to
$\ordsim0.8$ within the laser focus using the known fragmentation ratios of $\HHO^{+}$ and $\HHHOp$,
while the fragmentation channels of H$^{+}$, O$^{+}$, OH$^{+}$ have not been included. Taking the
known fragmentation channels into account, the fraction of the water dimer within the molecular beam
was evaluated. Comparing the ratios between the water dimer and all other species visible in the
TOF, a water dimer fraction of $3.9(6)~\%$ in the center of the undeflected beam and of $93(15)~\%$
in the deflected beam, at a position of +3~mm, was achieved. Thus, using the electrostatic deflector
the fraction of the water dimer within the interaction region could be increased by nearly a factor
of 24.

\section{Conclusions}
In summary, a high-purity beam of water dimers was created using the electrostatic deflector, which
spatially separated water dimers from other species present in the molecular beam. The resulting
water dimer sample had a purity of $93(15)~\%$. The fragmentation products and ratios of the water
dimer following strong-field ionization using a 35~fs laser pulse with a wavelength centered around
800~nm and peak intensity of $\ordsim10^{14}~\Wpcmcm$ were studied, with $46(4)~\%$ of the water
dimer found to form a hydronium ion and $46(7)~\%$ fragmenting into one water cation and one neutral
water monomer, while $8(2)~\%$ of the water dimer stayed intact. The deflection profiles could be
simulated using a rigid-rotor model and an initial rotational temperature of 1.5(5)~K.

The produced clean samples of water dimers are well suited for non-species-specific experiments,
\eg, reactive-collisions, diffractive imaging, or ultrafast
spectroscopies~\cite{Kilaj:NatComm9:2096, Hensley:PRL109:133202, Kuepper:PRL112:083002}. Even for
experiments that can distinguish different species, for example photoion-photoion coincidence
measurements~\cite{Ren:NatPhys:79:1745, Kierspel:PCCP20:20205}, the produced clean beams will enable
significantly faster measurements of this important hydrogen-bonded model system, \eg, because
unwanted backgrounds are avoided. Furthermore, the electrostatic separation technique can be used to
separate different conformers~\cite{Chang:IRPC34:557}, which could be highly interesting in the
purification and studies of larger water-clusters that exhibit multiple
conformers~\cite{Gregory:Science275:814}.

\section*{Acknowledgments}
\label{sec:Acknowledgments}
This work has been supported by the Clusters of Excellence ``Center for Ultrafast Imaging'' (CUI,
EXC~1074, ID~194651731) and ``Advanced Imaging of Matter'' (AIM, EXC~2056, ID~390715994) of the
Deutsche Forschungsgemeinschaft (DFG), by the European Union's Horizon 2020 research and innovation
program under the Marie Skłodowska-Curie Grant Agreement 641789 ``Molecular Electron Dynamics
investigated by Intense Fields and Attosecond Pulses'' (MEDEA), by the European Research Council
under the European Union's Seventh Framework Program (FP7/2007-2013) through the Consolidator Grant
COMOTION (ERC-Küpper-614507), and by the Helmholtz Gemeinschaft through the ``Impuls- und
Vernetzungsfond''. L.H.\ acknowledges a fellowship within the framework of the Helmholtz-OCPC
postdoctoral exchange program and J.O.\ gratefully acknowledges a fellowship by the Alexander von
Humboldt Foundation.

\section*{Supporting Information Description}
\label{sec:SupportingInformation}
Supporting Information Available: Description of the fragmentation correction method and the
trajectory simulations

\bibliography{string,cmi}%
\onecolumngrid
\end{document}


\title{Supplemental Material: Pure molecular beam of water dimer}%
\author{Helen Bieker}\cfeldesy\uhhphys\uhhcui%
\author{Jolijn Onvlee}\cfeldesy%
\author{Melby Johny}\cfeldesy\uhhphys%
\author{Lanhai He}\jilinu\cfeldesy%
\author{Thomas~Kierspel}\unibasel\cfeldesy\uhhphys\uhhcui%
\author{Sebastian Trippel}\cfeldesy\uhhphys%
\author{Daniel A. Horke}\uninijmegen\cfeldesy\uhhphys%
\author{Jochen Küpper}\jkemail\cmiweb\cfeldesy\uhhphys\uhhcui%
\maketitle

\section{Fragmentation correction of measurements}
The strong-field-ionization technique employed in this work can lead to fragmentation, such that
clusters from the molecular beam contributed to smaller masses in the mass spectrum (MS). For
example, the water monomer and the water dimer signals at $m/q=18$~u/e and 36~u/e, respectively,
contained contributions due to fragmentation of larger water-clusters in the molecular beam.
Therefore, measured intensities needed to be corrected for these fragmentation channels. In
addition, background water inside the chamber was measured at 18~u/e and needed to be corrected for.

For the latter, background measurement were permanently performed during the experiments using the
higher repetition rate of the laser compared to the valve. Laser pulses were arriving in the
interaction region at the same time as the molecular beam and between two molecular beam pulses,
such that for each data point a background measurement was performed. The background signal was
subtracted from the measurements.

\begin{figure}
   \includegraphics[width=\linewidth]{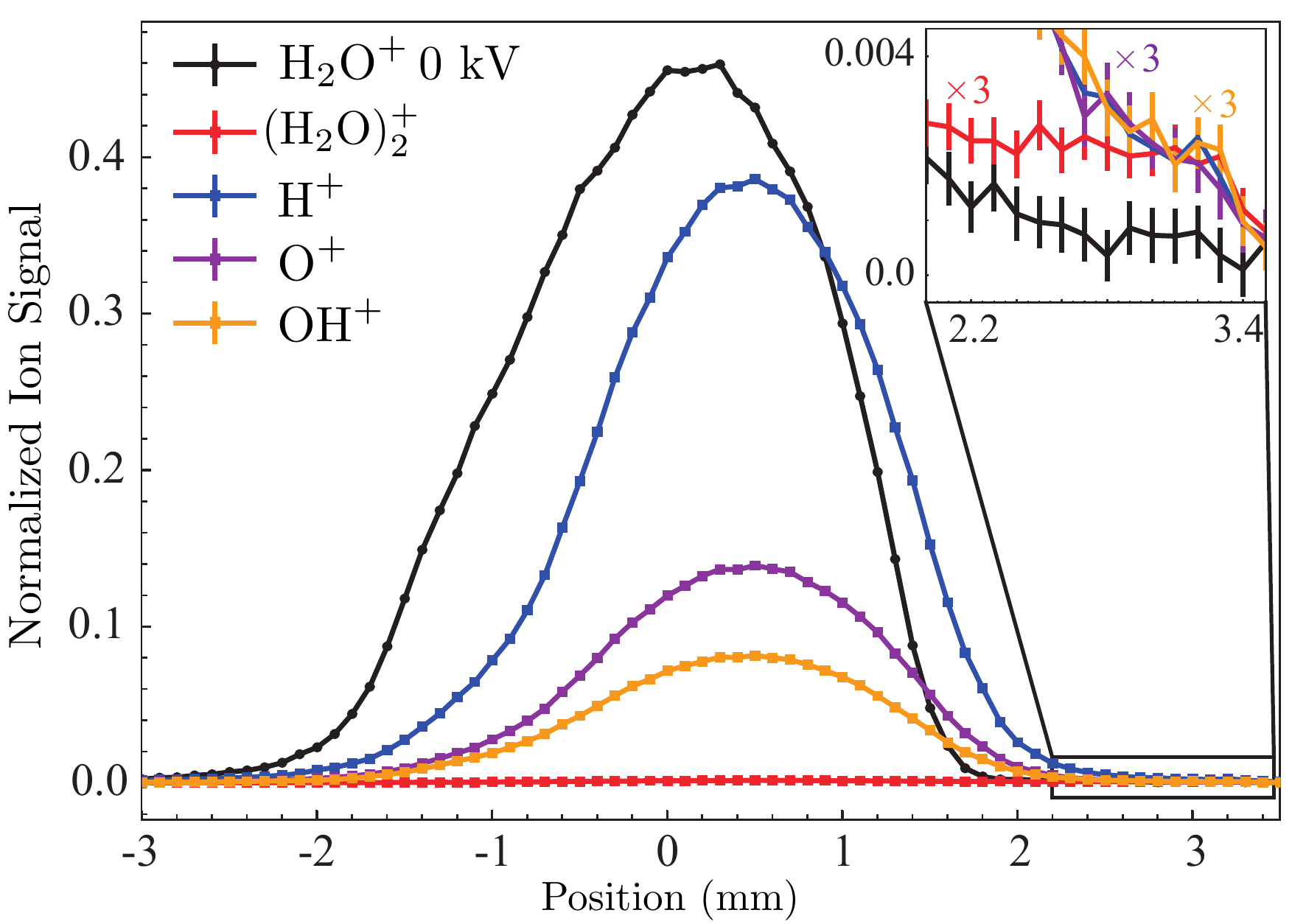}
   \caption{Column density profiles, measured for the water-monomer cation with deflector voltages
      of 0~kV (black) and for H$^+$ (blue), O$^+$ (purple), OH$^+$ (orange) and the water-dimer cation with
    a deflector voltage of 8~kV (red). The inset shows the region around $y=3$~mm enlarged, with
      O$^+$, OH$^+$ and the water dimer ion scaled by a factor $3$.}
   \label{fig:SI:deflection}
\end{figure}

The fragmentation ratios of the water dimer into smaller masses could be estimated and used for the
calculation of the fraction of the water dimer in the deflected and undeflected molecular
beam~\cite{Johny:CPL721:149}. In \autoref{fig:SI:deflection} the deflection profiles measured at
masses corresponding to H$^+$, O$^+$ and OH$^+$ are shown. In the region of 2.8 to 3.5~mm the
deflection curves look identical to those for the water dimer, indicating that at these positions
those are fragments from the water dimer. The ratios of the water dimer to H$^+$, O$^+$ and OH$^+$
at a position of 3~mm are 0.3, 0.8, and 0.7, respectively. For the calculation of the fraction of
the water dimer in the molecular beam for the undeflected beam, these ratios were used to estimate
the amount of the water dimer inside of the beam.

\begin{figure}
   \includegraphics[width=\linewidth]{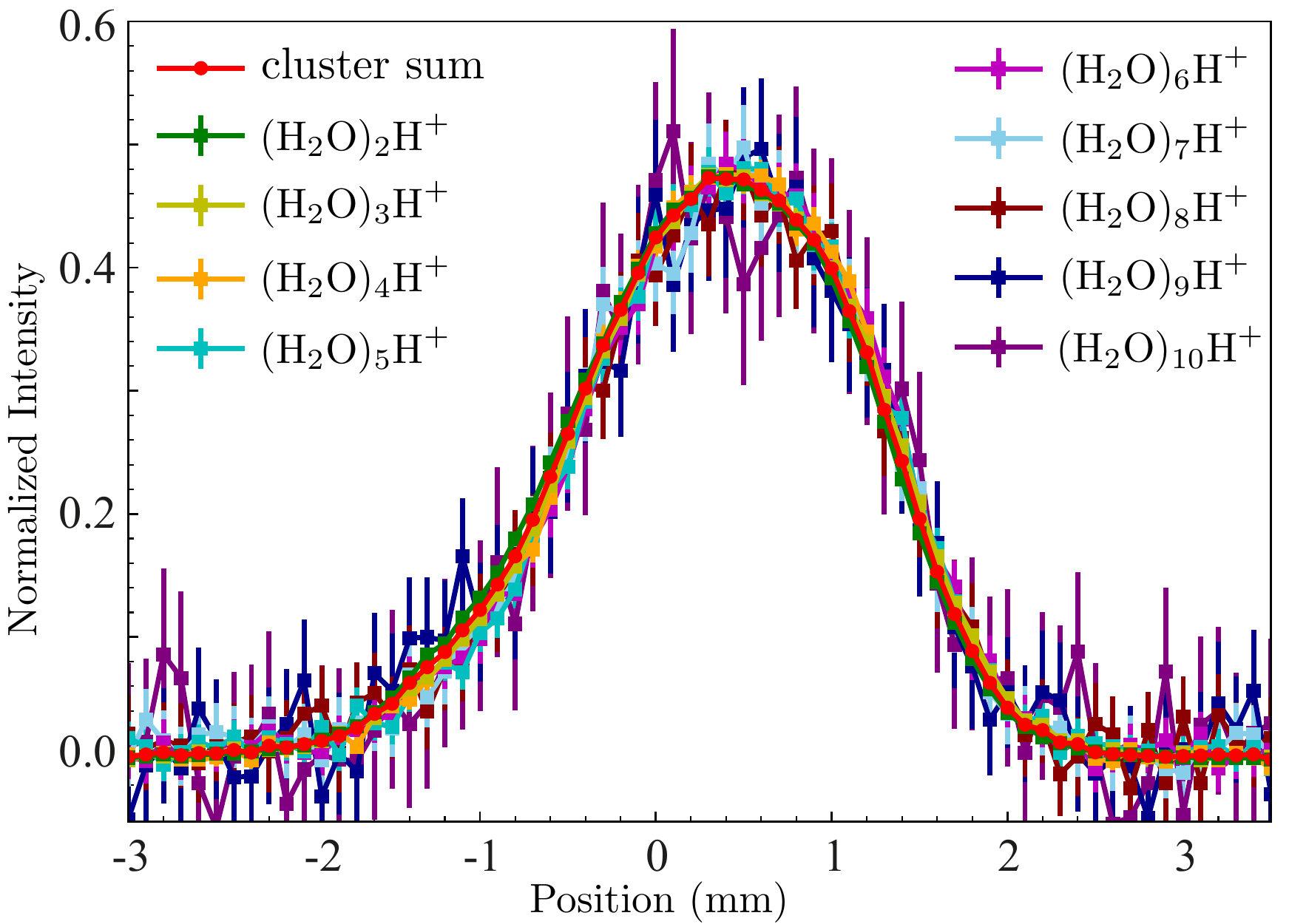}
   \caption{Comparison of the averaged measured protonated-water-cluster signal (red) and the
      individual measured protonated water-clusters deflection profiles for $n=2\ldots7$. The
      profiles are normalized to the area under the curve.}
   \label{fig:cluster_correction}
\end{figure}

For larger clusters, only fragments were measured, such that the measured signal was not solely due
to a specific cluster stoichiometry and the overall shape of the molecular beam profile arose from
several larger water-clusters. All protonated-water-cluster ions recorded showed the same deflection
behavior, see \autoref{fig:cluster_correction}. An estimate of the exponential decay of the measured
protonated water-clusters distribution showed that protonated water-clusters $n=1-10$ contained
$99.6~\%$ of the overall intensity.

\section{Trajectory Simulations}
\label{sec:trajectory_sim}
The Stark energies and effective dipole moments \mueff of water-clusters $n=1\ldots7$ were
calculated using the freely available \textsc{CMIstark} software package~\cite{Chang:CPC185:339},
which were then used to perform trajectory simulations~\cite{Filsinger:JCP131:064309} and to verify
the measured deflection profiles of water-clusters. The rotational constants, dipole moments and
centrifugal distortion constants from the literature are summarized in \autoref{tab:watercluster}.
\begin{table*}
   \scriptsize
   \begin{tabular}{l@{\hspace{0.5em}}d{3}d{3}d{3}l@{\hspace{0.5em}}d{2}d{2}d{2}l@{\hspace{0.5em}}d{6}d{6}d{6}d{6}d{7}l}
     \hline \hline
     molecule & \multicolumn{4}{c}{dipole moment $\mu$ (D)} & \multicolumn{4}{c}{rotational constants (MHz)} & \multicolumn{6}{l}{Centrifugal Distortion constants (kHz)}\\
              & \mu_a & \mu_{b}& \mu_{c} & & A & B & C & & \Delta_{J} & \Delta_{JK} & \Delta_{K} & d_{J}& d_{K}&  \\
     \hline
     $\HHO$ &0&  -1.86& 0 & \cite{Shostak:JCP94:5875}& 835840.29& 435351.72& 278138.70& \cite{DeLucia:JPCR3:211} &3.759\times10^{4}& -1.729\times10^{5}&9.733\times10^{5}& 1.521\times10^{4} & 4.105\times10^{4}&\cite{DeLucia:JPCR3:211}  \\
     $(\HHO)_2$&2.63& 0& 0 & \cite{Malomuzh:RussJPhysChemA88:1431}&190327.0& 6162.76& 6133.74 & \cite{Coudert:JMolSpec139:259}& 0.044& 4010& 0& 0& 0 &\cite{Dyke:JCP66:1977} \\
     $(\HHO)_3$ &0&0&0 & \cite{Gregory:Science275:814} &6646.91& 6646.91& 0 & \cite{Keutsch:ChemRev103:2533}& -&-&-&-&-& \\
     $(\HHO)_4$ &0& 0& 0  & \cite{Gregory:Science275:814} &3149.00& 3149.00& 1622.00& \cite{Cruzan:Science271:59} & -&-&-&-&- & \\
     $(\HHO)_5$&0.93& 0& 0 & \cite{Gregory:Science275:814}&1859.00& 1818.00& 940.00& \cite{Liu:Science271:62}& -&-&-&-&-& \\
     $(\HHO)_6$ book &0.17& 2.46&0.16 & \cite{Perez:Science336:897}&1879.47& 1063.98& 775.06& \cite{Perez:Science336:897}& -&-&-&-&-& \\
     $(\HHO)_6$ cage &1.63& 0.32& 1.13 & \cite{Perez:Science336:897} &2163.61& 1131.2& 1068.80 & \cite{Liu:Nature381:501} & -&-&-&-&-& \\
     $(\HHO)_6$ prism &2.41& 0.88& 0.42  & \cite{Perez:Science336:897}&1658.22& 1362.00& 1313.12 & \cite{Perez:Science336:897}& -&-&-&-&-& \\
     $(\HHO)_7$ 1 &1.0& 1.0& 0.0& \cite{Perez:ChemPhysLett571:1}&1304.44& 937.88& 919.52 & \cite{Perez:ChemPhysLett571:1}& 0.457& -0.342& 0.842& 0.0377& 0.63 &\cite{Perez:ChemPhysLett571:1} \\
     $(\HHO)_7$ 2 &1.0& 0.0& 1.0 & \cite{Perez:ChemPhysLett571:1}&1345.16& 976.88& 854.47 & \cite{Perez:ChemPhysLett571:1}&0.044& 0.000& 0.000& 0.0000497& 0& \cite{Perez:ChemPhysLett571:1}\\
     \hline \hline
   \end{tabular}
   \caption{Dipole moments, rotational constants and centrifugal distortion constants of
      water-clusters used in the Stark-effect calculations}
   \label{tab:watercluster}
\end{table*}
Three conformers for the water hexamer in prism-, book- and cage-like
form~\cite{Perez:Science336:897} and two conformers of the water heptamer following the naming
scheme of~\cite{Perez:ChemPhysLett571:1} were included.

For these simulations the water-clusters were assumed to be rigid rotors. Since the water dimer is
known to be a floppy molecule with large amplitude motions~\cite{Coudert:JMolSpec139:259,
   Odutola:JCP72:1980}, the corresponding energy spectra and the description of the interaction of
the states would significantly complicate further analysis. Using a rigid rotor assumption enables
an easier and faster description and it has been shown previously that this model can be used to
describe the dynamics of \indolew in strong-electric- and laser-field alignment and orientation
experiments~\cite{Trippel:PRA86:033202, Thesing:PRA98:053412} and to fit pure rotational transitions
of the water dimer to experimental measurements~\cite{Dyke:JCP66:1977}.

\begin{figure}
   \includegraphics[width=\linewidth]{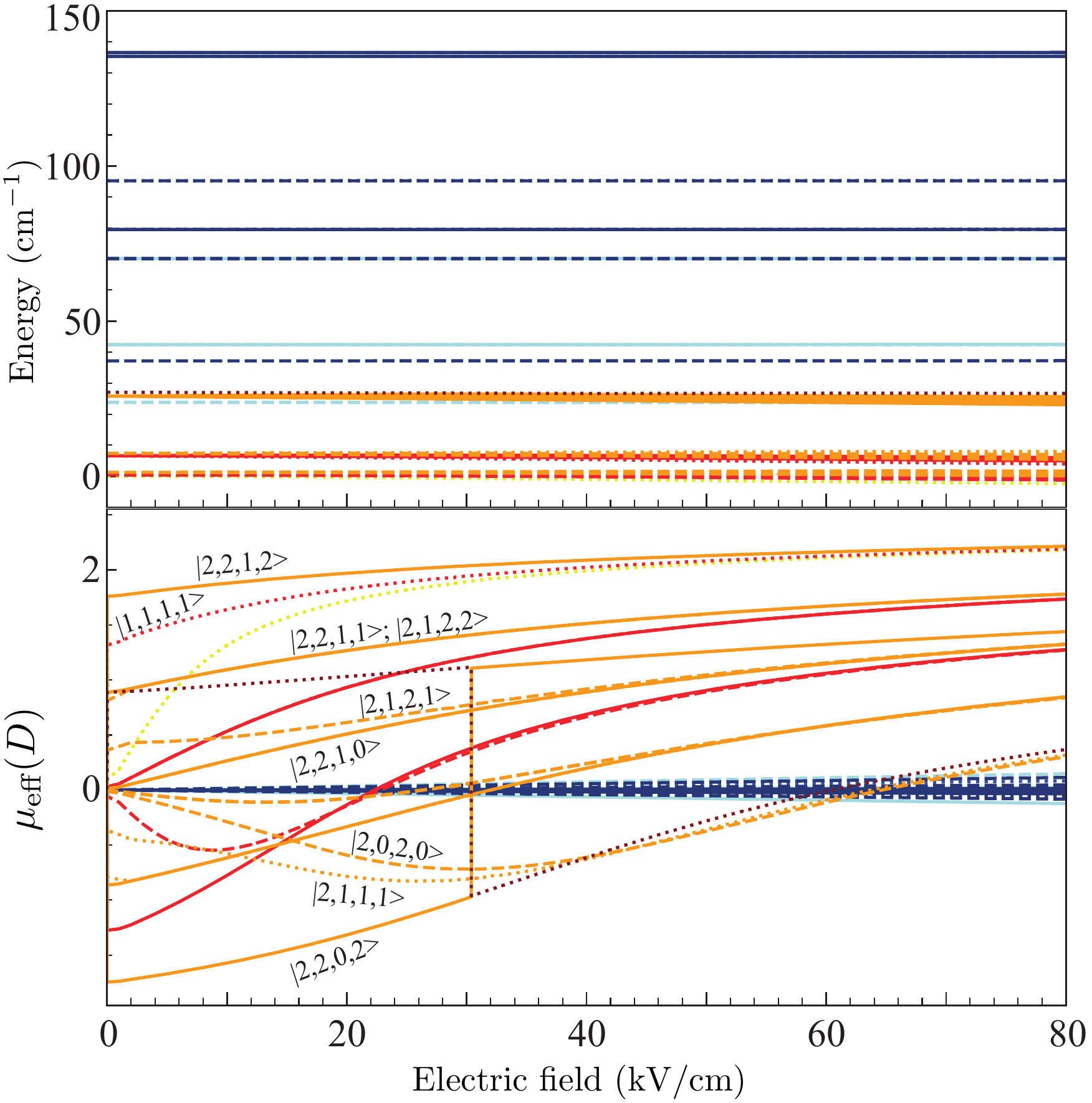}
   \caption{Calculated Stark energy and effective dipole moments for the $J=0\ldots2$ states of the
      water monomer (blue) and the water dimer (red, orange, yellow). $J=0$ are shown in blue
      (water) and yellow (water dimer), $J=1$ in light blue (water), red (water dimer), $J=2$ in
      darkblue (water) and orange (water dimer) and the \ket{3,2,2,2} state in darkred for the water
      dimer. States where $K_a<K_c$ are indicated by dashed lines, $K_a>K_c$ by solid lines and
      $K_a=K_c$ by dotted lines.}
   \label{fig:structure+energy}
\end{figure}

For the rotational states $J=0\ldots2$ of the water monomer and the water dimer the Stark energies and the
corresponding \mueff as a function of the electric field strength are shown in
\autoref{fig:structure+energy}. For the water dimer all relevant states are strong-field seeking and,
hence, accelerated toward regions of stronger fields. For a nominal field strength of 50~\kVpcm the
\mueff of the water dimer are significantly larger than for the water monomer, except from the
$\ket{J,K_a,K_c,M}=\ket{2,0,2,0},\ket{2,1,1,1}$ states, leading to a larger acceleration in the
electric field. All the shown states have a small asymmetry splitting, see \autoref{tab:watercluster},
resulting in a fast rise of $\mueff$ at small electric field strength. The discontinuous change of \mueff
at an electric field around 30~\kVpcm is ascribed to an avoided crossing of the \ket{2,2,0,2} and
\ket{3,2,2,2} states.

The trajectories of the molecules inside the electrostatic deflector were simulated using the
calculated \mueff~\cite{Filsinger:JCP131:064309}. For quantum states $J=0\ldots10$,
$10^{7}~$trajectories were calculated for each set of $J$ states and used to simulate the spatial
profiles using a weighting factor based on the thermal distributions of the state for a given
temperature. Those temperature-weighted simulated vertical molecular-beam profiles were scaled to
the area under the curve of the corresponding experimental profile to compare the deflection
profiles. The simulations include the nuclear-spin-statistical weights for the water monomer and the
water dimer. For \emph{para}- and \emph{ortho}-water a room-temperature distribution of $1:3$ was
used. The water dimer in its equilibrium geometry is isomorphic with the permutation-inversion point
group $D_{4h}$ including tunneling splittings~\cite{Dyke:JCP66:492}. Neglecting tunneling splittings
and acceptor switching, the rigid water dimer belongs to the symmetry group $C_S(M)$, yielding
nuclear-spin-statistical weights of \emph{para}$:$\emph{ortho} of
$16:16$~\cite{Bunker:FundamentalsMolecularSymmetry}.

\begin{figure}
   \includegraphics[width=\linewidth]{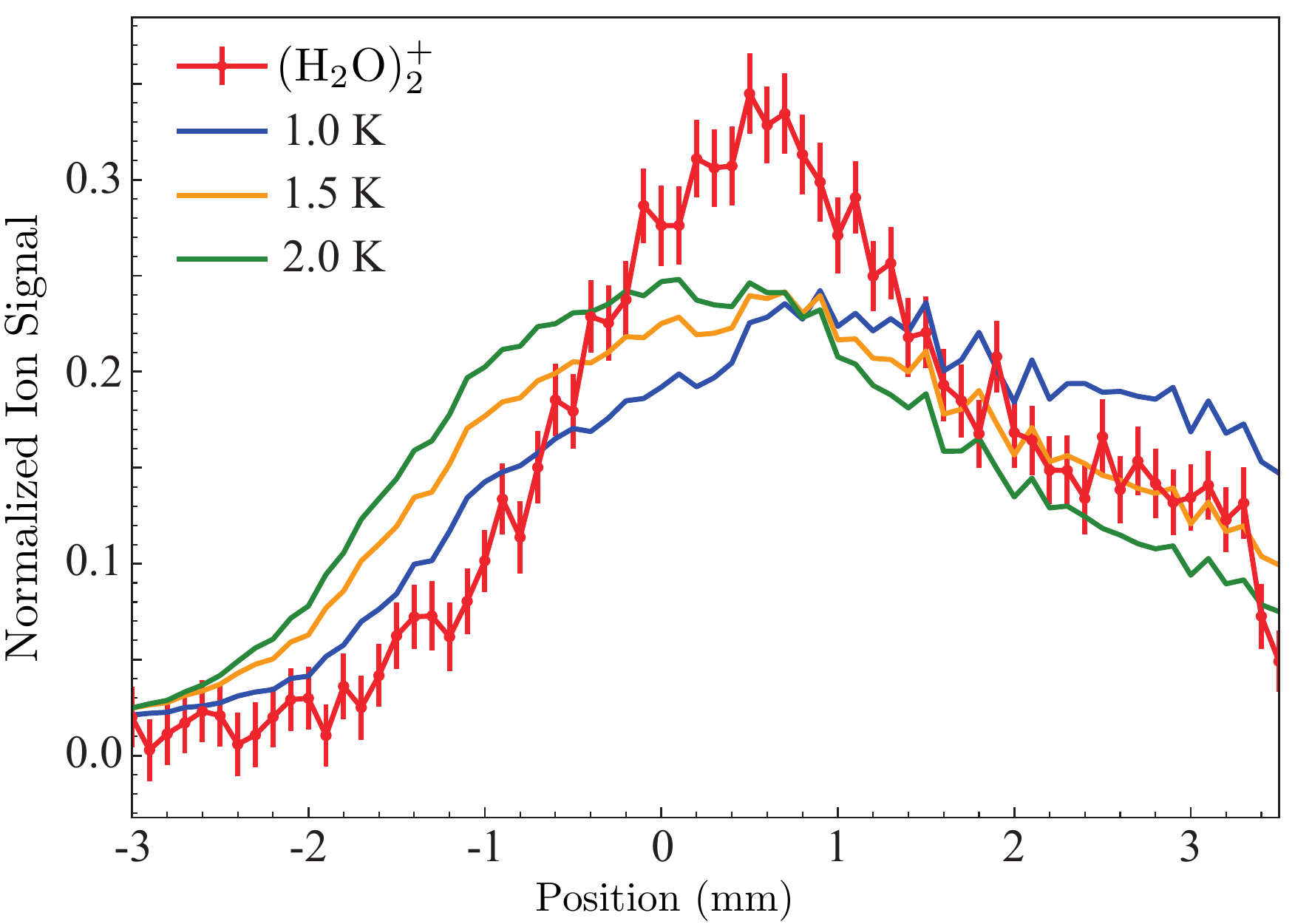}
   \caption{Simulated deflection profiles for the water dimer at temperatures of 1.0~K, 1.5~K and 2~K
      compared to the corrected pure water dimer profile at 8~kV (red, dots).}
   \label{fig:temp_fit_sim_dimer}
\end{figure}

The simulated profiles for the water dimer at different rotational temperatures $T_\text{rot}$ including
rotational states $J=0\ldots10$ are shown in \autoref{fig:temp_fit_sim_dimer}. An initial-beam
temperature of $T_\text{rot}=1.5(5)$~K reproduced the experiment the best. At this temperature
the water monomer in the para nuclear spin state has $100~\%$ of its population in its absolute rotational
ground states \ket{J=0,K_a=0,K_c=0,M=0}, while ortho-water populates the \ket{J=1,K_a=0,K_c=1,M=0,1}
state to equal amounts. $99.9~\%$ of the para-water dimer and $99.9~\%$ of the ortho-water dimer
population is within $J=0\ldots10$.

\begin{figure}
   \includegraphics[width=\linewidth]{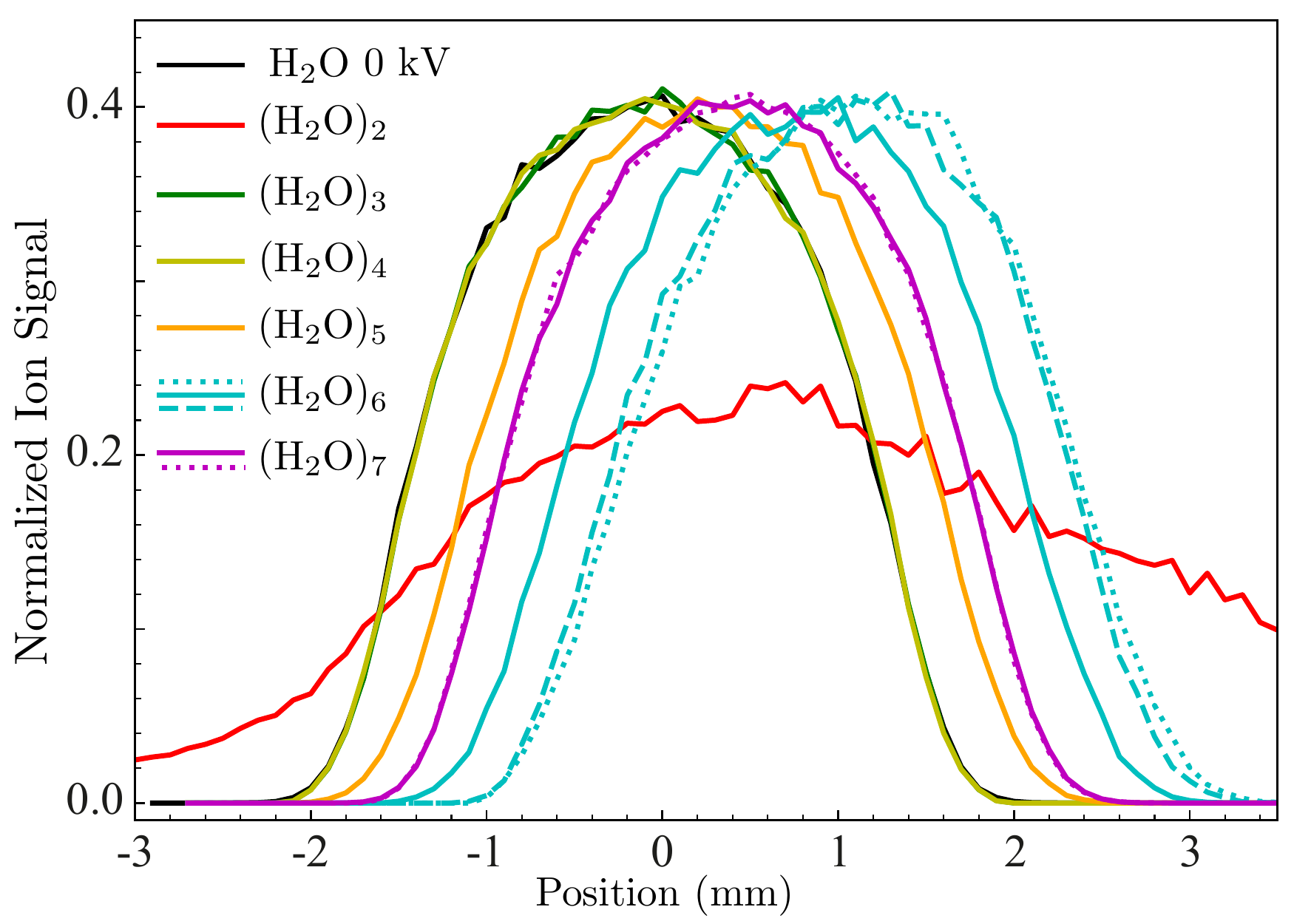}
   \caption{Simulated deflection profiles for water-clusters for $n=2\ldots7$ are shown. Three
      different conformers of the water hexamer and two of the water heptamer were included, see text
      for details.}
   \label{fig:simulation_clusters}
\end{figure}

Trajectory simulations were performed for water-clusters up to $n=7$. Based on the estimated
water-cluster distribution, \emph{vide supra}, this covers $97.8~\%$ of the water-clusters in the
molecular beam, while $\ordsim2.2~\%$ of the molecules in the beam are from water-clusters $n\ge8$.
The simulations for water-clusters including $J=0\ldots2$ and using the same rotational temperature
$T_\text{rot}=1.5(5)$~K of the water dimer are shown in \autoref{fig:simulation_clusters}.
We note that at this temperature rotational states up to $J=10$ might be populated in the molecular
beam and the rotational
temperature can differ from the one of the water dimer. Thus the simulations give just an estimate of
the amount of deflection. Based on the simulations the water dimer is deflecting the most of all
water-clusters, followed by the water hexamer in prism- and book-like form, which reaches to a
position of +3.2~mm.

\begin{figure}
   \includegraphics[width=\linewidth]{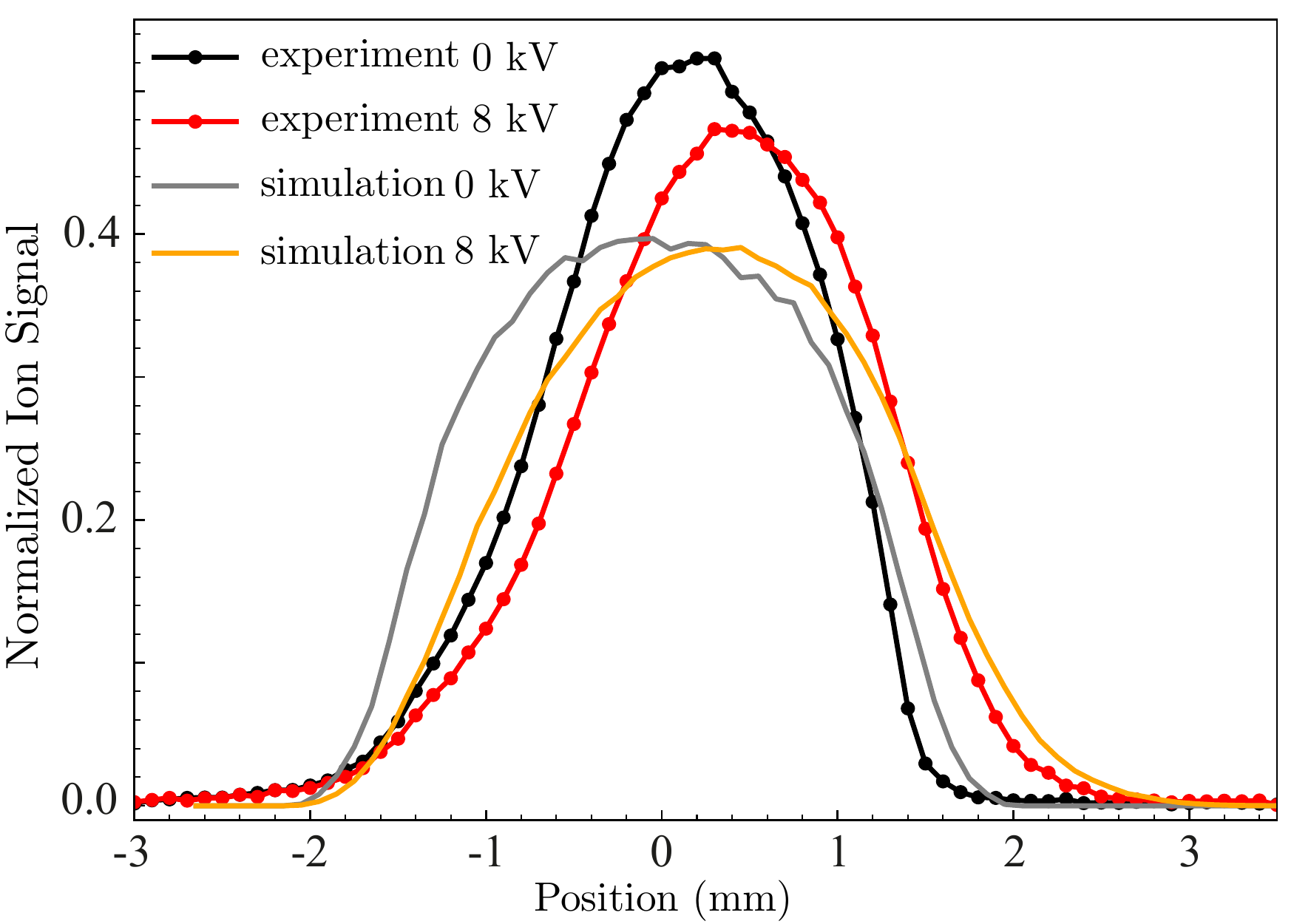}
   \caption{Summed up simulated deflection profiles for water-clusters for $n=3\ldots7$ (grey, orange)
      and summed up measured deflection profiles for protonated water-clusters for $n=2\ldots10$ (black
      and red dots) for a deflector voltage of 0~kV (grey, black) and 8~kV (orange, red). The
      profiles are normalized to the area under the curve.}
   \label{fig:simulation_clusters_09K}
\end{figure}

Since for larger clusters only fragments have been measured and, therefore, the shape of the
recorded beam profiles is the result of a superposition of several neutral cluster distributions in
the molecular beam, it is not possible to compare the single deflection profiles directly with
simulations. Therefore, at each position of the deflection profile the signal of the measured
protonated water-clusters for $n=2\ldots10$ have been summed up. The \HHHOp, $n=1$, contained also
signal
from the water dimer and has not been included. For the computationally derived profiles $n=3\ldots7$
were summed up for each position. As for the hexamer and heptamer several conformers have been
simulated, each profile of the hexamer has been divided by 3 and for the heptamer by 2. This is
shown in \autoref{fig:simulation_clusters_09K}.
These simulations assume a rather low temperature of 1.5~K and did not include the
needed nuclear spin statistical weighting for larger-clusters. In addition, the decaying water-cluster
distribution
in the molecular beam was not taken into account, resulting in a slightly different shape of the vertical
molecular-beam-density profiles than the measured ones. However, comparing the simulated and the
measured deflection profiles, the deflection is on the same order of magnitude and the right-hand side tail of
the
simulated deflection profile is reaching up to a position of +3.2~mm.

\bibliography{string,cmi}
\onecolumngrid